\documentstyle[aps,prl,twocolumn,epsf,floats]{revtex}

\newcommand{\beq}{\begin{equation}}
\newcommand{\eeq}{\end{equation}}
\newcommand{\bea}{\begin{eqnarray}}
\newcommand{\eea}{\end{eqnarray}}
\newcommand{\eps}{\epsilon}
\newcommand{\veps}{\varepsilon}
\newcommand{\etal}{{\it et al.},}
\newcommand{\nn}{\nonumber}

\begin{document}

\title{\bf \LARGE
The Vortex State in a Strongly Coupled Dilute
Atomic Fermionic Superfluid
}
\vspace{0.75cm}
\author{ Aurel Bulgac and Yongle Yu }
\vspace{0.50cm}

\address{Department of Physics, University of
Washington, Seattle, WA 98195--1560, USA}

\maketitle

\begin{abstract}

We show that in a dilute fermionic superfluid, when the Fermions
interact with an infinite scattering length, a vortex state is
characterized by a strong density depletion along the vortex core.
This feature can make a direct visualization of vortices in Fermionic
superfluids possible.

\end{abstract}

\draft


%
%
%
%


The existence of stable vortex states is one of the most spectacular
manifestations of superfluidity in both Bose and Fermi systems. In
Bose dilute atomic gases isolated quantized vortices \cite{madison}
and subsequently large arrays of vortices \cite{mit} have been
observed already. The quest for the observation of superfluidity in
dilute Fermi gases started as soon as the first experimental evidence
of an atomic dilute Fermi degenerate gas was published \cite{jin}.  In
Bose systems methods for establishing unambigously both the
superfluidity and the quantization of vortices exists \cite{nobel},
while most of the suggestions made so far for Fermi systems are rather
indirect. The
presence of vortices in BEC was confirmed by analyzing the density
variations of an expanding BEC cloud \cite{mit,nobel}. 
The Bose-Einstein condensation (BEC) creation was confirmed by 
studying the character of
the expansion of the atomic cloud after the trap was removed
\cite{becexp}. A recent experimental result in Fermi systems
\cite{dfbcs} suggests that a hydrodynamic expansion
of a Fermi superfluid is a plausible scenario \cite{expansion}.

In Fermi systems significant density variations due to the presence of
vortices are not expected \cite{nicolai}. We have reasons to expect
however, that under certain conditions the density variations induced
by the presence of one or more vortices could be akin to those
in the case of BEC, as a similar feature was recently established in
the analysis of the spatial structure of a vortex in low density
nuclear matter \cite{nmvortex}. At densities significantly smaller
than nuclear saturation densities, the superfluid gap in homogeneous
neutron matter can attain values rather large by normal standards,
$\Delta \approx 0.25 \veps_F$, where $\Delta$ is the value of the gap
and $\veps_F$ is the Fermi energy. Even though the gap is still
smaller than the Fermi energy, such values proved sufficient in the
case of low density neutron matter to lead to major density depletions
in the vortex core. As Pitaevskii and Stringari note \cite{dfbcs}, the
observation of quantized vortices in a dilute Fermi gas would provide
the ultimate proof that the system has undergone a transition to a
superfluid state. {\em 
A vortex is just about the only 
phenomenon in which a true stable superflow is created in a neutral 
system.} Other phenomena would only somewhat indirectly be affected 
by the onset of superfluidity.

Recently, a new {\it ab initio} calculation of the properties of low
density nuclear matter became available \cite{carlson1}. A particular
result of this analysis, which is valid in principle for any Fermion
system, concerns the properties of a two fermion species interacting
with an infinite scattering length. These authors have shown that the
energy per particle of such a normal Fermi system is
\beq
{\cal{E}}_N=\alpha_N \frac{3}{5}\veps_F, \label{eq:en}
\eeq
with $\alpha_N \approx 0.54$. This result was obtained for a system of
fermions interacting with a short range attractive potential with a
zero--energy bound state. As long as $k_Fr_0\ll 1$, where $r_0$ is of
the order of the radius of the potential, one expects on general
grounds that the energy per particle of such a system is proportional
to the energy per particle of a non--interacting system. The
scattering length $a$ and the effective range $r_0$ parameterize the
low energy behavior of the $s$--wave scattering phase of two particles
$k\cot \delta (k) \approx -1/a + r_0k^2/2,$
where $k$ is the wave vector of the relative motion. A second result
of this {\it ab initio} calculation concerns the energy per particle
of the superfluid phase of such a system \cite{carlson2}, namely that 
\beq
{\cal{E}}_S=\alpha_S  \frac{3}{5} \veps_F, \label{eq:es}
\eeq
with $\alpha_S \approx 0.44$. A simple estimate of the corresponding
value of the superfluid gap, using for the condensation energy the
weak coupling BCS value $-3\Delta^2/8\veps_F$ leads to $\Delta \approx
0.4 \veps_F$.  Such large values for the gap $\Delta$ are not
compatible with the BCS weak coupling limit, when $k_F|a|\ll 1$ and
$a<0$. It is well known that in this limit the BCS approximation
\cite{mohit} leads to a too large value of the gap and that
polarization corrections lead to a reduction of the gap
\cite{gorkov}, namely to
$$
\Delta
= \left (\frac{2}{e}\right )^{7/3}  \!\!\!\!\!
\veps_F \exp\left (  \frac{\pi}{2k_Fa} \right ) .
$$
A recent analysis \cite{nicolai} of the vortex state in a dilute
superfluid Fermi gas, using the simple BCS value for the gap (which
exceeds by a factor of $\approx 2.2$ the true gap value) shows a
relatively modest density depletion in the vortex core of about 10\% 
at most.

The possibility that the value of the superfluid gap can attain large
values was raised more than two decades ago in connection with the BCS
$\rightarrow$ BEC crossover \cite{leggett,nozieres}. One can imagine
that one can increase the strength of the two--particle interaction in
such a manner that at some point a real two--bound state forms, and in
that case $a\rightarrow -\infty$. By continuing to increase the
strength of the two--particle interaction, the scattering length
becomes positive and starts decreasing. When $a\gg r_0$ and $a>0$ the
energy of the two--particle bound state is
$\eps_2\approx -\hbar^2/ma^2.$
A dilute system of fermions, when $nr_0^3\ll 1$, will thus undergo a
transition from a weakly coupled BCS system, when $a<0$ and $a
={\cal{O}}(r_0)$, to a BEC system of tightly bound Fermion pairs, when
$a>0$ and $a ={\cal{O}}(r_0)$ again. In the weakly coupled BCS limit
the size of the Cooper pair is given by the so called coherence length
$\xi \propto \frac{\hbar^2k_F}{m\Delta},$
which is much larger than the interparticle separation $\approx
\lambda_F=2\pi/ k_F$. In the opposite limit, when $k_Fa\ll 1$ and
$a>0$, and when tightly bound pairs/dimers of size $a$ are formed, the
dimers are widely separated from one another. Surprisingly, these
dimers also repel each other with an estimated scattering length
$\approx 0.6...2a$ \cite{mohit,pieri} and thus the BEC phase is also
(meta)stable. The bulk of the theoretical analysis in the intermediate
region where $k_F|a|>1$ was based on the BCS formalism
\cite{mohit,leggett,nozieres,randeria,rest} and thus is highly
questionable. All these authors have considered only the simple ladder
diagrams in the particle--particle channel. The inclusion of the
additional "bosonic" degrees of freedom \cite{rest}, which represent
nothing else but a two--atom bound state, falls into the same
approximation scheme, which includes only ladder diagrams in the
atom--atom channel. Even the simplest polarization corrections have
not been included into this type of analysis so far. In particular, it
is well known that in the low density region, where $a<0$ and
$k_F|a|\ll 1$ the polarization corrections to the BCS equations lead
to a noticeable reduction of the gap \cite{gorkov}.  Only a truly {\it
ab initio} calculation could really describe the structure of a many
Fermion system with $k_F|a|\gg 1$. In the limit $a=\pm \infty$, when
the two--body bound state has exactly zero energy, and if $k_Fr_0\ll
1$, one can expect that the energy per particle of the system is
proportional to $\veps_F=\hbar^2k_F^2/2m$, as it was recently
confirmed by the variational calculations of
Refs. \cite{carlson1,carlson2}.

As in any other Fermion systems \cite{kohn}, the knowledge of the
energy per particle as a function of density in the homogeneous phase,
Eq. (\ref{eq:en}), allows us to construct the normal part of the Local
Density Approximation (LDA) for the energy density functional
(EDF). The extension of the LDA to superfluid systems presented in
Refs. \cite{lda} in conjunction with Eq. (\ref{eq:es}) lead us to a
rather well defined EDF, appropriate for Fermion systems with infinite
scattering length, namely
\bea
& & {\cal{E}}(\bbox{r})n (\bbox{r}) =
\frac{\hbar^2}{m}\left [
\frac{m}{2m^*}\tau(\bbox{r}) +
\beta n (\bbox{r}) ^{5/3} +
\gamma \frac{|\nu(\bbox{r})|^2}{n(\bbox{r})^{1/3}}\right ],\label{eq:ed}\\
& & n (\bbox{r}) =\sum_\alpha |v_\alpha (\bbox{r})|^2,\quad
 \tau (\bbox{r}) =\sum_\alpha |\bbox{\nabla}v_\alpha (\bbox{r})|^2, \nn \\
& & \nu (\bbox{r}) =\sum_\alpha v_\alpha^* (\bbox{r})u_\alpha (\bbox{r}). \nn
\eea
Here $n (\bbox{r})$ and $\nu (\bbox{r})$ are the normal and anomalous
densities and $\tau (\bbox{r})$ is the kinetic energy density, all of
them expressed through the quasiparticle wave functions
$(u_\alpha(\bbox{r}), v_\alpha (\bbox{r}))$.  The summation over
$\alpha$ should be interpreted as a sum or integral when appropriate,
over all quasiparticle states with the Bogoliubov--de Gennes
eigenvalues $E_\alpha>0$.  The dimensionless couplings $m/m^*$,
$\beta$ and $\gamma$ have to be chosen so as to reproduce in the case
of infinite homogeneous matter the expressions
(\ref{eq:en}--\ref{eq:es}) for the energy per particle. Unfortunately
the parameterization of the EDF is not unique and its form has to be
restricted using some additional considerations. One possible choice
corresponds to $m/m^*=\alpha_N$, $\beta=0$ and $\gamma\approx
-6.64$. We shall refer to this case as the parameter set I. Another
possible choice of parameters, referred to as the parameter set II,
corresponds to $m/m^*=1$, $\beta = 3(\alpha_N -1)(3\pi^2)^{2/3}/5$ and
$\gamma \approx -7.20 $. Since the anomalous density $\nu(\bbox{r})$
is a diverging quantity, a regularization procedure of the above
expression (\ref{eq:ed}) is required. This was performed according to
the formalism described in great detail by us elsewhere
\cite{lda}. The renormalization procedure amounts to replacing
$\gamma$ with a new well defined running coupling
$\gamma_{\mathit{eff}}(\bbox{r})$. Even though in this renormalization
procedure there is an explicit energy cutoff appearing in the
formalism, no observable is affected by its presence, when this energy
cutoff is chosen appropriately.  Up to the overall factor $\hbar^2/m$,
the energy density in Eq. (\ref{eq:ed}) has the overall scaling
${\cal{E}}(\bbox{r})\propto [n(\bbox{r})]^{5/3}$, which is expected
from dimensional arguments for a system with an infinite scattering
length and $k_Fr_0\ll 1$. In principle, there are an infinite number
of possible local energy densities satisfying this requirement.  In
particular, one could have considered for the superfluid part of the
functional for example a contribution of the form $\propto
|\nu(\bbox{r})|^{5/3}$. Since a term $\propto |\nu(\bbox{r})|^2$ works
in both BCS and BEC limits and since the emerging pairing field
$\Delta (\bbox{r})$ has the simplest form and the renormalization
procedure is also the simplest when using the EDF Eq. (\ref{eq:ed}),
we did not consider explicitly any other forms. Moreover, nobody ever
suggested other forms for the contribution of the pairing correlations
to the energy density in the past, as far as we are aware. We do not
expect however any qualitative changes in our results resulting from
considering other forms for the superfluid contribution to the energy
density.

\begin{figure}[tbh]
\begin{center}
\epsfxsize=6.0cm
\centerline{\epsffile{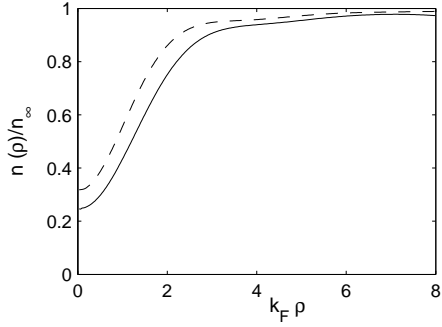}}
\end{center}
\caption{ The density profile of a vortex with the symmetry axis $Oz$
as a function of $k_F\rho $, where $k_F$ is the value of the Fermi
wave vector in the homogeneous phase $n_{\infty}$ is the asymptotic
value of the density. The solid line is for the parameter set I, while
the dashed line is for the set II.  }

\label{fig:fig1}
\end{figure}

\begin{figure}[tbh]
\begin{center}
\epsfxsize=6.0cm
\centerline{\epsffile{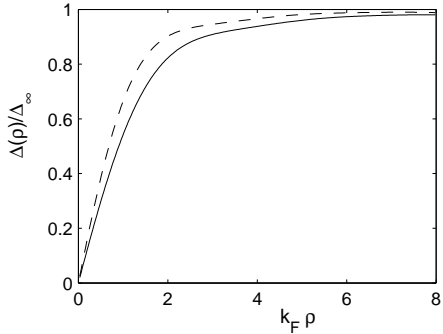}}
\end{center}
\caption{ The pairing field profile, where $\Delta_{\infty}$ is the
asymptotic value of the pairing field and the rest of the notations
are identical to those in Fig. \ref{fig:fig1}.  }

\label{fig:fig2}
\end{figure}
\begin{figure}[tbh]
\begin{center}
\epsfxsize=6.0cm
\centerline{\epsffile{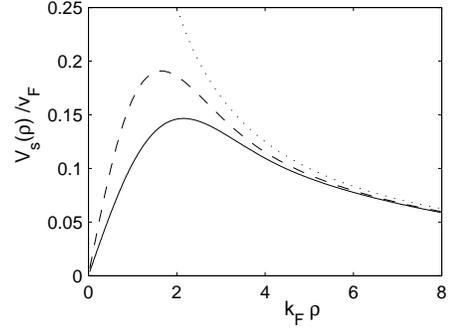}}
\end{center}
\caption{ The velocity profile in units of the Fermi velocity in the
homogeneous phase ${\mathrm{v}}_F= \hbar k_F/m$ and the rest of the
notations are identical to those in Fig. \ref{fig:fig1}. The dotted
line corresponds to the ideal vortex velocity profile $V_v(\rho ) =
\hbar /2m\rho$.  }

\label{fig:fig3}
\end{figure}

We looked for a self--consistent solution of the Bogoliubov--de Gennes
equations corresponding to the EDF in Rel. (\ref{eq:ed}), with a
vortex in the pairing field along the $Oz$--axis and $\hbar/2$ of net
angular momentum per particle, namely $\Delta (\bbox{r}) = \Delta
(\rho ) \exp ( i \phi )$, where $\bbox{r} =(\rho, \phi, z)$ are
cylindrical coordinates.  The mean--field depends only on the radial
coordinate $\rho$.  The quasiparticle wave functions have the
structure
\beq
\left (
\begin{array}{l}
u_\alpha(\bbox{r})\\ v_\alpha (\bbox{r})
\end{array}
\right )
=\left (
\begin{array}{l}
u_s(\rho )\exp\left [ in\phi +  ikz\right ] \\
v_s(\rho )\exp\left [ i(n-1)\phi + ikz\right ]
\end{array}
\right ), \nn
\eeq
where $s$ labels the radial part of the quasiparticle states, $n$ is
an integer, $k$ is the wave vector of the quasiparticle state along
the vortex symmetry axis $Oz$ and $\alpha=(s,k,n)$.  The presence of
the vortex implies a net (super)flow of the Fermion superfluid around the
vortex axis with the velocity profile
\beq
{\mathrm{V}}_s(\rho )\bbox{\hat{e}_\phi}
=\frac{\hbar}{m\rho}\frac{1}{n(\rho )}\sum_{\alpha}
v^*_{\alpha} (\bbox{r})i
\bbox{\hat{e}_\phi}\frac{\partial}{\partial \phi}
v_{\alpha}(\bbox{r}), \nn
\eeq
where $\bbox{\hat{e}_\phi} =(y,-x,0)/\rho $,

The most salient feature emerging from this analysis is the behavior
of the density as a function of the radial coordinate $\rho$, the
prominent density depletion on the vortex axis, see
Fig. \ref{fig:fig1}.  The density at the vortex core is lowered to a
value of  $\approx 0.3 \rho_\infty$, while the pairing field
vanishes at the vortex axis as expected, see Fig. 2.  
By comparing the actual flow profile with that of an ideal vortex, 
see Fig. 3, one can see that only the fraction 
$f_s(\rho ) = {\mathrm{V}}_s(\rho )/ V_v(\rho )$ of the system is superfluid
at a distance $\rho$ from the vortex axis.
We have performed also
calculations by imposing the artificial constraint that the pairing
field in the homogeneous phase has a larger value than the one
expected from the self--consistent solution, corresponding to the
EDF Eq. (\ref{eq:ed}). In that case the density
depletion at the vortex core becames even larger. In hindsight this
result could have been expected. Large values of the pairing field
correspond to the formation of atom pairs/dimers of relatively small
sizes. When these dimers are relatively strongly bound and when they
are also widely separated from one another, they undergo a
Bose--Einstein condensation. For a vortex state in a 100\% BEC system
the density at the vortex axis vanishes identically. Therefore, by
increasing the strength of the two--particle interaction, the Fermion
system simply approaches more and more an ideal BEC system, for which
a density depletion of the vortex core is expected.

The value of the coupling constant, and correspondingly the value of
the scattering length can nowadays be routinely controlled in
experiments \cite{feshbach_e}, following earlier theoretical
suggestions \cite{feshbach_t}. One can then expect that it would be
feasible to create a BCS state as in the experiment of
Ref. \cite{dfbcs}, stir the system so as to create one or several
vortices in a manner similar to the experiments with dilute Bose gases
\cite{madison,mit,nobel} and change the ambient magnetic field so that
the system is brought close to the Feshbach resonance. The first
quantized vortex will appear only at some critical angular velocity,
if the system is gradually spun up. The vortex quantization could be
demonstrated for example as in Ref. \cite{madison}.  At zero
temperature, the single vortex state is the lowest energy state with
the total angular momentum $N\hbar/2$ (here $N$ is the total number of
particles) and its decay into lower energy states with total angular
momenta differing by a few number of $\hbar/2$ is strongly suppressed.
The next vortex will appear at a larger critical angular velocity, and
so forth. The density profiles of the vortices now develop significant
density depletions at the cores, which could hopefully be subsequently
visualized, if they survive the expansion when the trap is removed. As
discussed in Ref. \cite{expansion}, a fluid with a power law equation
of state (the present case) will have a simple scaling expansion, see
also Refs. \cite{becexp,dum}. As a consequence, the density depletion
along the vortex core should survive the atomic cloud expansion upon
trap release.  Note that a system in which the scattering length is
infinite is highly unusual, as, even when it expands and its density
decreases, the relative role of the interaction does not change,
unlike any other interacting system, and such a system remains
strongly coupled at all densities.

By increasing further the strength of the interaction, two atoms will
form a bound state. In the limit when the size of this bound state
becomes significantly smaller then the mean interparticle separation
the system becomes a BEC of atom pairs. These pairs repel each other
with an estimated scattering length $\approx 0.6 \dots 2a$
\cite{mohit,pieri}. The BEC phase can in principle be described as
well within the framework of the same formalism outline above. Such an
approach would be unreasonable, since an accurate description of the
relatively strongly bound two--atom system in terms of extended
mean--field quasiparticle states would require an exceedingly large
number of quasiparticle states. The ratio of the quantum states
required for a reasonable accuracy to the number of particles per
particle would be at least of order $1/(na^3) \gg 1$.  The natural
description in this limit is in terms of bosonic dimers, which weakly
repel each other. The magnitude of the coherence length in the BEC
phase is exceeding considerably the mean separation between pairs,
$\xi_b \approx \sqrt{na} \gg 1/n^{1/3}$. This coherence length $\xi_b$
determines the radius of the vortex core. Thus along with the
two--atom scattering length $a$, the size of the vortex core could be
controlled as well by means of the Feshbach resonance.

In summary, we have shown that the vortex in a dilute atomic
superfluid Fermi gas interacting with an infinite scattering length
develops a significant density depletion along its core, which could
be visualized after expanding the atomic cloud.

We thank P.F. Bedaque, G.F. Bertsch, J. Carlson, E.K.U. Gross,
N. Hayashi, V.R. Pandharipande, D.S. Petrov, B. Spivak, 
S. Stringari and A. Svidzinsky for discussions and/or correspondence. This work
benefited from partial financial support under DOE contract
DE--FG03--97ER41014.



\begin{thebibliography}{99}


\bibitem{madison} K.W. Madison, F. Chevy, 
J. Mod. Opt. {\bf 47}, 2715 (2000);
F. Chevy, \etal Phys. Rev. Lett. {\bf 85}, 2223 (2000).

\bibitem{mit} J.R. Abo--Shaeer, \etal
Science {\bf 292}, 476 (2001).

\bibitem{jin} B. DeMarco and D.S. Jin, Science, {\bf 285}, 1703
(1999).

\bibitem{nobel} E. A. Cornell and C. E. Wieman, Rev. Mod. Phys. {\bf
74}, 875 (2002); W. Ketterle, Rev. Mod. Phys. {\bf 74}, 1131 (2002).

\bibitem{becexp} Yu. Kagan, \etal
Phys. Rev. A {\bf 54}, 1753(R) (1996); U. Ernst, \etal Appl. Phys. {\bf B 67},
719 (1998).

\bibitem{dfbcs} K.M. O'Hara, \etal
Science, {\bf 298}, 2179 (2002);
L. Pitaevskii and S. Stringari, Science, {\bf 298}, 2144 (2002).

\bibitem{expansion} C. Menotti, \etal
Phys. Rev. Lett. {\bf 89}, 250402 (2002).

\bibitem{nicolai} N. Nygaard, \etal
Phys. Rev. Lett. {\bf 90}, 210402 (2003); \O. Elgar{\o}y and
F.V. De Blasio, A\&A, {\bf 370}, 939 (2001).

\bibitem{nmvortex} Y. Yu and A. Bulgac, Phys. Rev. Lett. {\bf 90},
161101 (2003).

\bibitem{carlson1} J. Carlson, \etal
 nucl-th/0302041.

\bibitem{carlson2} J. Carlson, \etal
Phys. Rev. Lett. {\bf 91}, 050401 (2003).

\bibitem{mohit} M. Randeria, in {\it Bose--Einstein Condensation},
eds. A. Griffin, \etal Cambridge Univ.
Press (1995), pp 355--392.


\bibitem{gorkov} L.P. Gorkov and T.K. Melik--Barkhudarov,
Zh. Eksp. Teor. Fiz. {\bf 40}, 1452 (1961) [Sov. Phys. JETP {\bf 13},
1018 (1961)]; H. Heiselberg, \etal
Phys. Rev. Lett. {\bf 85}, 2418 (2000).

\bibitem{leggett} A.J. Leggett, in {\it Modern Trends in the Theory of
Condesed Matter}, eds. A. Pekalski and R. Przystawa, Springer--Verlag,
Berlin, 1980; J. Phys. (Paris) Colloq. {\bf 41}, C7--19 (1980).

\bibitem{nozieres} P. Nozi\`eres and S. Schmitt--Rink, J. Low
Temp. Phys. {\bf 59}, 195 (1985).

\bibitem{pieri} P. Pieri and G.C. Strinati, Phys. Rev. B {\bf 61},
15370 (2000) and cond-mat/0307421; D.S. Petrov, C. Salomon, 
G.V. Shlyapnikov,  cond-mat/0309010.

\bibitem{randeria} C.A.R. S\'a de Mello, \etal
Phys. Rev. Lett. {\bf 71}, 3202 (1993);
J.R. Engelbrecht, \etal Phys. Rev. B
{\bf 55}, 15153 (1997).

\bibitem{rest} Y. Ohashi and A. Griffin, Phys. Rev. Lett. {\bf 89},
130402 (2002) and earlier references therein.

\bibitem{kohn} P. Hohenberg and W. Kohn, Phys. Rev. {\bf 136}, B864
(1964); W. Kohn and L.J. Sham, Phys. Rev. {\bf 140}, A1133 (1965);
R.M. Dreizler and E.K.U. Gross, {\it Density Functional Theory: An
Approach to the Quantum Many--Body Problem}, (Springer, Berlin, 1990);
R.G. Parr and W. Yang, {\it Density--Functional Theory of Atoms and
 Molecules}, Clarendon Press, Oxford (1989).

\bibitem{lda} A. Bulgac and Y. Yu, Phys. Rev. Lett. {\bf 88}, 042504
(2002) and nucl-th/0109083;
A. Bulgac, Phys. Rev. C {\bf 65}, 051305(R) (2002);
Y. Yu and A. Bulgac, Phys. Rev. Lett. {\bf 90}, 222501 (2003).

\bibitem{feshbach_e} S. Inouye {\it et al}, Nature,.  {\bf 392}, 151
(1998); J.L. Roberts {\it et al}, Phys. Rev. Lett. {\bf 81}, 5109
(1998); J. Stenger {\it et al}, J. Low Temp. Phys. {\bf 113}, 167
(1998); P. Courteille {\it et al}, Phys. Rev. Lett. {\bf 81}, 69
(1998).

\bibitem{feshbach_t} E. Tiesinga {\it et al}, Phys. Rev. A {\bf 47},
4114 (1993); P.O. Fedichev {\it et al}, Phys. Rev. Lett. {\bf 77},
2913 (1996); J.L. Bohn and P.S. Julienne, Phys. Rev. A {\bf 56}, 1486
(1997); A.J. Moerdijk {\it et al}, Phys. Rev. A {\bf 53}, 4343 (1996);
J.M. Vogels {\it et al}, Phys. Rev. A {\bf 56}, R1067 (1997);
M. Marinescu and L. You, Phys. Rev. Lett. {\bf 81}, 4596 (1998).

\bibitem{dum} Y. Castin and R. Dum, Phys. Rev. Lett. {\bf 77}, 5315 (1996).

\end{thebibliography}
\end{document}